\documentclass[copyright,creativecommons,submission]{eptcs}
\providecommand{\event}{Programming Language Approaches to Concurrency and\\
Communication-cEntric Software 2026 (PLACES'26)}

\usepackage{iftex}

\ifpdf
  \usepackage[strings]{underscore}
  \usepackage[T1]{fontenc}
\else
  \usepackage{breakurl}
\fi

\usepackage{amsmath}
\usepackage{amssymb}
\usepackage{mathpartir}
\usepackage{stmaryrd}
\usepackage{setspace}
\usepackage{ebproof}
\usepackage{makecell}
\usepackage{graphicx}
\usepackage{xfrac}
\usepackage{xcolor}
\usepackage{subfig}
\usepackage{mathtools}
\usepackage{amsthm}
\usepackage{listings}
\usepackage{float}
\usepackage{titlesec}
\usepackage{enumitem}

\title{Branching Out: Existential External Choice in Effpi}

\author{Benjamin Robinson
\institute{University of Oxford, Oxford, UK}
\email{benjamin.robinson@st-hughs.ox.ac.uk}
\and
Nobuko Yoshida
\institute{University of Oxford, Oxford, UK}
\email{nobuko.yoshida@cs.ox.ac.uk}}
\def\titlerunning{Branching Out: Existential External Choice in Effpi}
\def\authorrunning{B. Robinson \& N. Yoshida}

\begin{document}
% Proof trees
% Expect previously: \def\ruleset{\Gamma} (for example)
\newcommand{\rulename}[1]{\text{\tiny[\ruleset-#1]}}
\newcommand{\andConclusion}{\hspace{1em}}
\ebproofnewstyle{small}{separation = 1em, rule margin = .5ex, template = \footnotesize$\inserttext$ }
\newcommand{\tree}[2][1em]{\begin{prooftree}[small]#2\end{prooftree}\hspace{#1}}
\newcommand{\colorTree}[2][1em]{\colorbox{yellow}{$\begin{prooftree}[small]#2\end{prooftree}$}\hspace{#1}}
\ebproofnewstyle{smallCoinductive}{separation = 1em, rule margin = .5ex, rule style = double, template = \footnotesize$\inserttext$ }
\newcommand{\coTree}[2][1em]{\begin{prooftree}[smallCoinductive]#2\end{prooftree}\hspace{#1}}
\newcommand{\colorCoTree}[2][1em]{\colorbox{yellow}{$\begin{prooftree}[smallCoinductive]#2\end{prooftree}$}\hspace{#1}}
\newcommand{\rulesMinipage}[1]{
    \setstretch{2.75}
    \begin{minipage}{13cm}
        \vspace{0.75em}
        \centering
        #1
        \vspace{0.75em}
    \end{minipage}
}
\newcommand{\semanticRuleName}[1]{\text{\tiny[T$\rightarrow$#1]}}

% Typing judgements
\newcommand{\judgeEnv}[1]{\vdash #1 \ \text{env}}
\newcommand{\judgeType}[2][\Gamma]{#1 \vdash #2 \ \text{type}}
\newcommand{\judgePiType}[2][\Gamma]{#1 \vdash #2 \ \pi\text{-type}}
\newcommand{\judgePiInType}[2][\Gamma]{#1 \vdash #2 \ \pi\text{-in-type}}
\newcommand{\judgeSubtype}[3][\Gamma]{#1 \vdash #2 \leq #3}
\newcommand{\judge}[3][\Gamma]{#1 \vdash #2 : #3}
\newcommand{\judgeStarType}[2][\Gamma]{#1 \vdash #2 \ \text{*-type}}

% Citations
\newcommand{\etal}{\textit{et al.}}

% Types
\newcommand{\bool}{\text{bool}}
\newcommand{\proc}{\textbf{proc}}
\newcommand{\nil}{\textbf{nil}}
\newcommand{\cin}[1]{\text{c}^\text{i}[#1]}
\newcommand{\cout}[1]{\text{c}^\text{o}[#1]}
\newcommand{\cio}[1]{\text{c}^\text{io}[#1]}
\newcommand{\oType}[3]{\textbf{o}[#1, #2, #3]}
\newcommand{\iType}[2]{\textbf{i}[#1, #2]}
\newcommand{\pType}[2]{\textbf{p}[#1, #2]}
\newcommand{\bType}[2]{\textbf{b}[#1, #2]}
\newcommand{\tType}[2]{\textbf{timed}[#1, #2]}
\newcommand{\labelledType}[2][\ell]{\langle #1 : #2 \rangle}

% Processes
\newcommand{\bools}{\mathbb{B}}
\newcommand{\variables}{\mathbb{X}}
\newcommand{\channels}{\mathbb{C}}
\newcommand{\terms}{\mathbb{T}}
\newcommand{\values}{\mathbb{V}}
\newcommand{\processes}{\mathbb{P}}
\newcommand{\labels}{\mathbb{L}}
\newcommand{\ifThenElse}[3]{\textbf{if}\ #1\ \textbf{then}\ #2\ \textbf{else}\ #3}
\newcommand{\letIn}[2]{\textbf{let}\ #1 = #2\ \textbf{in}\ }
\newcommand{\newChan}{\textbf{chan}()}
\newcommand{\err}{\textbf{err}}
\newcommand{\pEnd}{\textbf{end}}
\newcommand{\send}[3]{\textbf{send}(#1,\ #2,\ #3)}
\newcommand{\recv}[2]{\textbf{recv}(#1,\ #2)}
\newcommand{\branch}[2]{\textbf{branch}(#1,\ #2)}
\newcommand{\labelled}[2][]{\ell_{#1} (#2)}
\newcommand{\timeout}[2]{\textbf{timeout}(#1,\ #2)}

% Sets / indexing helpers
\newcommand{\indexedSet}[2]{\{ {#1}_{#2} \}_{#2 \in \MakeUppercase{#2}}}
\newcommand{\labelSet}[3][\ell]{\{ {#1}_{#3} \mapsto {#2} \}_{#3 \in \MakeUppercase{#3}}}
\newcommand{\indexedOr}[2]{\bigvee_{#2 \in \MakeUppercase{#2}} {#1}_{#2}}

% Theorem styles
\theoremstyle{definition} \newtheorem{theorem}{Theorem}[section]
\theoremstyle{definition} \newtheorem{example}{Example}[section]
\theoremstyle{definition} \newtheorem{contributions}{Contributions}[section]

% Code listings....
\lstdefinelanguage{scala}{
  morekeywords={abstract,case,catch,class,def,%
    do,else,extends,false,final,finally,%
    for,if,implicit,import,match,mixin,%
    new,null,object,override,package,%
    private,protected,requires,return,sealed,%
    super,this,throw,trait,true,try,%
    type,val,var,while,with,yield,%
    given,using},
  otherkeywords={=>,<-,<\%,<:<,<:,>:,\#,@,>>:,>>,=:=},
  sensitive=true,
  morecomment=[l]{//},
  morecomment=[n]{/*}{*/},
  morestring=[b]",
  morestring=[b]',
  morestring=[b]"""
}

\lstset{
  language=scala,
  basicstyle=\footnotesize\ttfamily,
  keywordstyle=\color{blue}\bfseries,
  commentstyle=\color{gray}\itshape,
  stringstyle=\color{red},
  numbers=left,
  numberstyle=\tiny\color{gray},
  stepnumber=1,
  numbersep=8pt,
  showstringspaces=false,
  breaklines=true,
  frame=single,
  tabsize=2,
  captionpos=b,
  xleftmargin=2em,
  framexleftmargin=1.5em
}

% Double angle brackets
\makeatletter
\newsavebox{\@brx}
\newcommand{\llangle}[1][]{\savebox{\@brx}{\(\m@th{#1\langle}\)}%
  \mathopen{\copy\@brx\kern-0.5\wd\@brx\usebox{\@brx}}}
\newcommand{\rrangle}[1][]{\savebox{\@brx}{\(\m@th{#1\rangle}\)}%
  \mathclose{\copy\@brx\kern-0.5\wd\@brx\usebox{\@brx}}}
\makeatother
\newcommand{\ccst}[1]{\llangle #1 \rrangle_{\Gamma}}

% Misc
\newcommand{\dom}[1]{\text{dom($#1$)}}
\newcommand{\dep}[2]{\Pi(\underline{#1}:#2)}
\newcommand{\substitution}[2]{\{\sfrac{#1}{#2}\}}
\newcommand{\sep}{\ \mid\ }
\newcommand{\reducesTo}{\longrightarrow}
\newcommand{\context}{\mathcal{E}}
\newcommand{\lambdaPi}{\lambda^\pi_\leq}
\newcommand{\extendedLambdaPi}{{\setlength{\fboxsep}{0pt}\colorbox{yellow}{\rule[-3pt]{0pt}{\dimexpr1em+3pt\relax}\kern1pt Extended\kern1pt}}-$\lambdaPi$}
\newcommand{\effpi}{\texttt{Effpi}}

\newcommand{\lts}[4][\Gamma]{#1 \vdash #2 \xrightharpoondown{\text{\tiny $#3$}} #4}
\newcommand{\ltsStar}[4][\Gamma]{#1 \vdash #2 \xrightharpoondown{\text{\tiny $#3$}}^{\raisebox{-1ex}{$\scriptstyle *$}} #4}
\newcommand{\typeTransition}[4][\Gamma]{#1 \vdash #2 \xrightarrow{\text{\tiny $#3$}} #4}
\newcommand{\typeTransitionStar}[4][\Gamma]{#1 \vdash #2 \xrightarrow{\text{\tiny $#3$}}^{\raisebox{-1ex}{$\scriptstyle *$}} #4}

% Paragraph spacing
% {Left margin}{space above}{space after}
\titlespacing*{\paragraph}{0pt}{1.0ex plus 1ex minus .2ex}{0.5em}

% List spacing
\setlist{itemsep=2pt, topsep=2pt}

% Reduce space before/after figures
\setlength{\intextsep}{6pt plus 2pt minus 2pt}
\setlength{\textfloatsep}{6pt plus 2pt minus 2pt}
\setlength{\floatsep}{6pt plus 2pt minus 2pt}
\maketitle

\begin{abstract}
    \effpi\ \cite{scalas_effpi_2019, scalas_verifying_2019} is a framework for
    writing strongly-typed message-passing programs in Scala, where the compiler
    enforces the conformance of process implementations to specified protocol
    types. A compiler plugin is provided to verify properties of protocols, such
    as deadlock-freedom and liveness, by encoding the behavioural types into a
    variant of CCS \cite{milner_communication_1989}.
    
    To address limitations in the expressiveness of the existing toolkit, we
    extend \effpi\ with external choice by introducing a branching operation.
    Upon accepting a message via a \texttt{branch}, protocols enforce a
    continuation which depends on the label (type) of the received message.
    We equip the branching operation with the ability to accept messages over
    more than one channel. Additionally, we introduce a ``catch timeout''
    operation to allow processes to gracefully handle a lack of incoming
    messages. The enhanced expressiveness of \effpi\ is demonstrated through
    a number of examples, including an implementation of the Raft consensus
    algorithm \cite{ongaro_raft_2014}.
\end{abstract}
\section{Introduction \& Motivation}

Concurrent programs are notoriously difficult to implement correctly. Nuanced
bugs often arise from unexpected interactions between numerous components
and are hard to reason about. Issues include deadlocks, livelocks, or simple
protocol violations. Where possible, detecting and preventing such
issues at compile-time is highly desirable. Behavioural type systems, such as
session types \cite{honda_session_1993}, provide a formalism for encoding a
protocol as a type, allowing for static conformance verification.

\effpi\ \cite{scalas_effpi_2019, scalas_verifying_2019} is a Scala 3 toolkit which aims to address
these challenges. The framework provides an embedded DSL for writing
message-passing programs, whose implementations are checked against
a specified protocol, written as a dependent function type
\cite{scala_dependent_function_types}. This allows channels to
be tracked at the type-level, which facilitates model checking: protocols (even
when defined as a type without a concrete implementation) can be
verified for properties such as deadlock-freedom and liveness \cite{scalas_verifying_2019}. A
compiler plugin encodes the behavioural types in a variant of CCS \cite{milner_communication_1989},
and subsequently uses mCRL2 \cite{groote_mcrl2_2007}.

Despite its many strengths, \effpi\ has some limitations in expressiveness,
which we now illustrate by means of two motivating examples. In both, we
consider a hypothetical protocol and attempt to encode it as a behavioural
type in the existing framework. We hope to represent the protocol faithfully
so that the type system can be used to fully verify the correctness of
the corresponding implementation.

\begin{example}
  Consider a travel agency process which receives an
  accept/reject message from a client and provides a ticket accordingly,
  inspired by \cite{yoshida_very_2020}. The protocol
  is represented by the \texttt{TravelAgency} type and implemented concretely
  by the \texttt{travelAgency} method. A key limitation is that the
  protocol does not enforce a ticket being returned in
  exactly the \texttt{Accept} case: line 5 uses the
  union type ``\texttt{|}'' to specify \textit{internal} choice between providing
  a ticket or not; in reality, this choice should be external (dependent only on the client's message).
  The concrete process could currently provide tickets
  to clients who rejected the offer and still technically conform to the type.
  Ideally, the protocol should force the implementation to
  send a ticket if and only if the client accepted.

  \begin{lstlisting}[language=Scala]
case class Accept(); case class Reject(); type Decision = Accept | Reject

type TravelAgency[C1 <: IChan[Decision], C2 <: OChan[String]] = 
  In[C1, Decision, (d: Decision) => // receive the client's decision
    Out[C2, String] | PNil] // send a ticket or stop

def travelAgency(c1: IChan[Decision], c2: OChan[String]):
  TravelAgency[c1.type, c2.type] = {
    recv(c1) { (decision: Decision) =>
      case Accept() => send(c2, "Your ticket")
      case Reject() => nil // sending a ticket would still type-check here!
    }
}
\end{lstlisting}

  \label{ex:travel-agency-faulty}
\end{example}

\begin{example}
  \label{ex:auction-house-faulty}

  Consider a hypothetical \textit{AuctionHouse} protocol, which accepts
  \textit{Bid} messages on one channel and \textit{CloseAuction} messages on
  another (for example, control messages from the auctioneer). Upon receiving a
  \textit{CloseAuction} message, suppose that the protocol requires no
  further \textit{Bid} messages be accepted. To implement this in the
  existing \effpi\ toolkit, we could try to merge the two original
  channels, so that a single \texttt{receive} operation can listen to both bids and control messages.
  We might attempt to use separate components to forward incoming messages
  onto the shared channel, as illustrated in Figure~\ref{fig:auction-house-issue}.
  However, this subtly adapts the protocol: upon closing the auction, a bid
  may get trapped in the forwarding process and
  therefore might never be delivered to the \texttt{AuctionHouse} (despite the
  client seeing the synchronous communication as successful).
  
  Suppose further that we expect the \texttt{AuctionHouse} process to respond to
  \textit{a lack of} incoming bids: for example, to lower the starting price.
  This cannot be achieved within the existing framework as there is no way
  to catch the timeout of a \texttt{receive} operation.\footnote{The naive evaluation
  strategy throws a global runtime exception on timeout, but this cannot be
  caught in situ. The more sophisticated runtime systems do not handle timeouts
  at all. Details can be found in the original \effpi\ codebase \cite{scalas_effpi_2019}.}
  As with the rest of the protocol, this could be implemented as a concrete process, but
  we cannot represent this behaviour at the type-level.
  
  \begin{figure}[h]
    \centering
    \vspace{0.3em}
    \includegraphics[width=0.75\textwidth]{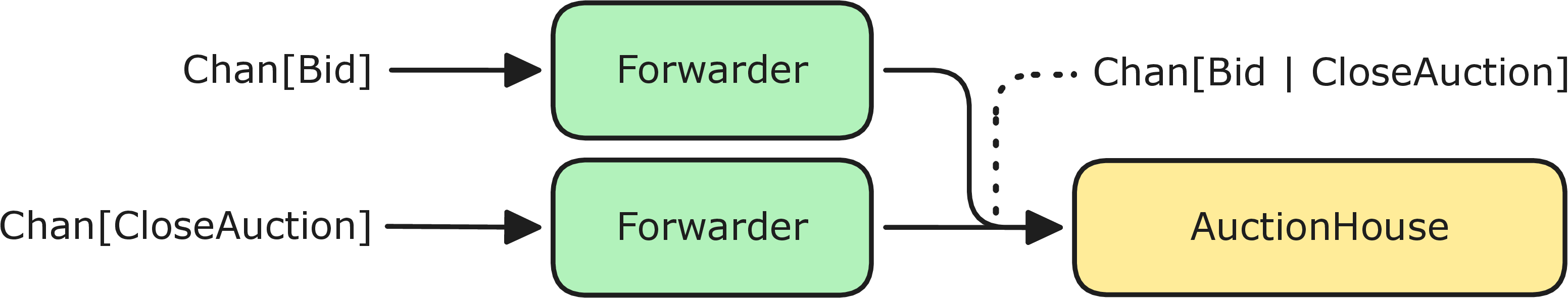}

    \caption{A faulty implementation of the \texttt{AuctionHouse} protocol by merging two channels.}
    \label{fig:auction-house-issue}
  \end{figure}
\end{example}

With these motivating examples, we have identified three limitations of the
existing toolkit: \textbf{the lack of external choice,} \textbf{the inability to
listen on multiple channels simultaneously,} and \textbf{the inability to handle
timeouts on inputs.} Whilst match types \cite{scala_match_types} were a promising initial direction
of research, they addressed only one of the above limitations and became
verbose for large protocols.\footnote{Due to the necessity of
subtyping \texttt{Process}, each match type would need a separate \texttt{type}
declaration.} Instead, we introduce a
custom branching operation, over possibly multiple input channels, and a
``catch timeout'' operation. We present the design of this
extension to the embedded DSL and demonstrate its expressiveness through an
implementation of the Raft consensus algorithm \cite{ongaro_raft_2014}. In
doing so, we discover further ways in which the framework
could be improved. The modified \effpi\ software toolkit and examples are available at:

\begin{center}
  \def\UrlFont{\bfseries}
  \url{https://github.com/benrobinson16/effpi}
\end{center}
\vspace{0.25em}

\section{Formal Calculus \& Type System}\label{sec:calculus}

% FOR SHORT VERSION:
The \effpi\ toolkit is built on the formal model of the $\lambdaPi$-calculus
and its type system \cite{scalas_verifying_2019}. To extend the toolkit
with new operations, we present an extended calculus, which we call
\extendedLambdaPi. We briefly outline possible modifications to the
type system and semantics in the full version of this paper \cite{full_version}.
The focus remains on the practical implementation and applicability
of an extension to \effpi, with formal soundness proofs of the extended type
system left to future work.

% FOR LONG VERSION:
% The \effpi\ toolkit is built on the formal model of the $\lambdaPi$-calculus
% and its type system \cite{scalas_verifying_2019}. To extend the toolkit
% with new operations, we present an extended calculus, which we call
% \extendedLambdaPi, and provisionally outline the corresponding modifications
% to the type system and semantics (provided in full in the appendices).
% The focus remains on the practical implementation and applicability
% of an extension to \effpi, with formal soundness proofs of the extended type
% system left to future work.

The extended syntax is shown in Figure \ref{fig:calculus-syntax},
with the new constructs highlighted. We introduce a set of labels
$\labels$ which are used to tag values.
The \textbf{branch} operation takes a set of channels to listen on
and a set of labelled continuations: upon receiving a labelled message on any
of its channels, the corresponding continuation is executed, matching the label received. Meanwhile, the
\textbf{timeout} operation is modelled on CSP's sliding-choice operator
\cite{roscoe_tpc_2000}: $\timeout{t}{t'}$ behaves initially as $t$
before becoming $t'$ if no input is received. At the formal level,
we refrain from specifying the exact timing of the timeout, instead
treating it as a non-deterministic tau action.

\begin{figure}[h]
    \footnotesize
    \setstretch{1.4}
    \[
        \begin{gathered}
            \bools = \{ \textbf{tt},\ \textbf{ff} \} \qquad
            \channels = \{ \text{a},\ \text{b},\ \text{c},\ \dots \} \qquad
            \variables = \{ x,\ y,\ z,\ \dots \} \quad
            \colorbox{yellow}{$\labels = \{ \ell_1,\ \ell_2,\ \ell_3,\ \dots \}$}
            \\
            \begin{array}{@{}rcl@{}}
                \text{terms}\ \terms \owns t, t', \dots
                & ::=
                & \variables
                    \sep \values
                    \sep \neg t
                    \sep \ifThenElse{t}{t_1}{t_2}
                    \sep \letIn{x}{t} t'
                    \sep t\ t'
                    \sep \newChan
                    \sep \processes
                \\
                \text{values}\ \values \owns u, v, \dots
                & ::=
                & \bools
                    \sep \channels
                    \sep \lambda x.t
                    \sep ()
                    \sep \err
                    \sep \colorbox{yellow}{$\labelled{t}$}
                \\
                \text{processes}\ \processes \owns p, q, \dots
                & ::=
                & \pEnd
                    \sep \send{t}{t'}{t''}
                    \sep \recv{t}{t'}
                    \sep t \parallel t'
                    \sep \colorbox{yellow}{$\branch{\indexedSet{t}{j}}{\labelSet{t_k}{k}}$}
                \\
                &
                & \colorbox{yellow}{$\timeout{t}{t'}$}
                    \hfill \text{where}\ J,K \text{ are finite, non-empty index sets}
            \end{array}
        \end{gathered}
    \]
    \caption{Syntax of the \extendedLambdaPi\ calculus.}
    \label{fig:calculus-syntax}
\end{figure}

The syntax of types is extended similarly, as shown in Figure
\ref{fig:type-syntax}. We add support for unions over index sets, which we assume
to be finite and non-empty. The indexed union can be
viewed as a generalisation and replacement of the previous binary type
$T \lor U$, and is useful when representing the possible input values to a branching
operation (by taking the union over the continuation labels and arguments). The
type $\langle \ell : T \rangle$ represents a labelled value,
and the types $\bType{...}{...}$ and $\tType{...}{...}$ represent the branching
and timeout operations respectively.

\begin{figure}[h]
    \footnotesize
    \setstretch{1.4}
    \[
        \begin{gathered}
            \bool
                \sep ()
                \sep \top
                \sep \bot
                \sep \dep{x}{U} T
                \sep \mu \underline{x}.T
                \sep \underline{x}
                \sep \colorbox{yellow}{$\indexedOr{T}{j}$}
                \sep \colorbox{yellow}{$\labelledType{T}$}
                \sep \cio{T}
                \sep \cin{T}
                \sep \cout{T}
            \\
                \proc
                \sep \nil
                \sep \oType{S}{T}{U}
                \sep \iType{S}{T}
                \sep \pType{T}{U}
                \sep \colorbox{yellow}{$\bType{\indexedSet{S}{j}}{\labelSet{U_k}{k}}$}
                \sep \colorbox{yellow}{$\tType{T}{U}$}
        \end{gathered}
    \]
    \caption{Syntax of types for \extendedLambdaPi.}
    \label{fig:type-syntax}
\end{figure}

% FOR SHORT VERSION:
Extended typing rules and semantics must build on the existing rules
from \cite{scalas_verifying_2019}. As part of these rules, we expect to
enforce conditions on the new constructs: for the timeout operation,
we require that the first term is either a receive or a branch
(the operations that can time out) and the second term is a lambda
abstraction. For branching, we require:

% FOR LONG VERSION:
% We propose extended typing rules in Appendix \ref{appendix:type-system},
% building on the existing rules from \cite{scalas_verifying_2019}. As part of these rules, we expect to
% enforce conditions on the new constructs: for the timeout operation,
% we require that the first term is either a receive or a branch
% (the operations that can time out) and the second term is a lambda
% abstraction. For branching, we require:

\begin{itemize}
    \item \textbf{Label distinctness:} Each continuation must have a distinct label.
    \item \textbf{Continuation coverage:} Every possible incoming label has a continuation.
    \item \textbf{Payload compatibility:} The payload type of each incoming message
    must be compatible with the expected argument type of the corresponding continuation.
\end{itemize}

\begin{example}\label{ex:travel-agency-calculus}
    Consider again the travel agency protocol from Example \ref{ex:travel-agency-faulty}.
    In the extended type system, we represent the protocol more accurately by
    using the branching type: the expected continuations after
    receiving accept/reject labels are given explicitly. We define
    $T_{\text{Decision}} = \labelledType[\ell_{\text{Accept}}]{()} \lor \labelledType[\ell_{\text{Reject}}]{()}$
    to be the union of two labelled unit types, representing the client's decision.
    \begin{align*}
        T_{\text{Agency}} = \dep{c1}{\cin{T_{\text{Decision}}}}\ \dep{c2}{\cout{string}}\ \textbf{b}\left[\{\underline{c1}\},
        \left\{
        \begin{array}{@{}l@{}}
            \ell_{\text{Accept}} \mapsto \Pi \lparen\rparen\ \oType{\underline{c2}}{string}{\Pi \lparen\rparen\ \nil } \\
            \ell_{\text{Reject}} \mapsto \Pi \lparen\rparen\ \nil
        \end{array}
        \right\}
        \right]
    \end{align*}
    A corresponding process is given below. Using extended typing rules,
    we verify that the process correctly implements the protocol; more
    precisely, we derive $\judge{myAgency}{T_{\text{Agency}}}$.
    No process which returns a ticket after
    a $\ell_{\text{Reject}}$ label can be typed with $T_{\text{Agency}}$, so we have
    improved implementation safety.
    \begin{align*}
        myAgency = \lambda c1 .\ \lambda c2 .\ \textbf{branch}\left(
        \{c1\},                                                                                                       
        \left\{                                                                                             
        \begin{array}{@{}l@{}}                                                                                                                                         
            \ell_{\text{Accept}} \mapsto \lambda \_ .\ \send{c2}{\text{``ticket''}}{\lambda \_ .\ \pEnd } \\                                                                  
            \ell_{\text{Reject}} \mapsto \lambda \_ .\ \pEnd
        \end{array}
        \right\}
        \right)
    \end{align*}
\end{example}
\section{Extending the Effpi Toolkit}

\subsection{Existing Codebase}

\effpi\ is implemented as an embedded DSL. At its core
are classes representing the possible actions of a process, such as
sending and receiving. Functions are provided to instantiate
these classes in a manner that appears natural: as though the operation is
truly happening at the point of invocation. Instead, the produced
hierarchy of classes and continuations is interpreted at runtime.
The framework additionally includes a compiler plugin, which encodes the
behavioural types for model checking. The plugin is not
modified in this paper, and it is left for future work to extend it for the
new operations.

\subsection{DSL Design}

\paragraph{Branching.} Like the existing
constructs in \effpi, we introduce the branching operation by extending
\texttt{Process}. We use tuples to represent the set of channels and
continuations, as this allows for variable length
whilst maintaining the specific type information of each element.
Using a list would cast the channels and continuations to a common supertype,
losing the vital type information required to enforce correct implementations.
The specific type information is also required for model checking later on.

\begin{lstlisting}[language=Scala]
case class Branch[A, Chans <: Tuple, Matches <: Tuple]
  (channels: Chans, matches: Matches, timeout: Duration)
  (using val valid: ValidBranch[A, Chans, Matches],
         val wrapper: WrapMatches[A, Matches]) extends Process
\end{lstlisting}

Implementing this construct has a few challenges; notably, we must enforce
the validity of the operation (as previously defined) and also maintain the type information of
each of the continuation arguments. To enforce the operation's validity:

\begin{itemize}
  \item \textbf{Label distinctness:} In the formal model, we require each
  continuation label to be distinct. In Scala, we use types (e.g. case classes)
  as labels, but checking for overlapping types is non-trivial. One approach is
  to require sealed traits
  and use \texttt{Mirror.SumOf[A]} to analyse distinctness. However, the
  loss of flexibility when using overlapping label sets throughout a larger
  program is undesirable. Instead, we re-interpret the implementation as
  imposing an ordering on the continuations (which is natural for a
  \texttt{Tuple}). At runtime, the first matching continuation is selected.

  \item \textbf{Continuation coverage:} Every incoming message
  (every instance of \texttt{A}) must have a corresponding continuation:
  or more precisely, the union of continuation argument types must cover \texttt{A}.
  This is checked via a \texttt{using} clause \cite{scala_using},
  which requires an implicit instance of \texttt{ValidBranch} to be witness to
  the validity of the branching operation. This instance is provided by a
  \texttt{given} definition, listed
  below, allowing it be automatically derived at compile-time so the DSL
  code remains clear \cite{scala_givens}.

  \item \textbf{Payload compatibility:} The provided channels must support only
  subtypes of \texttt{A}. Again, this is checked via the \texttt{ValidBranch}
  implicit instance.
\end{itemize}

\begin{lstlisting}[language=Scala]
given valid[A, Chans <: Tuple, Matches <: Tuple](using
  // The channels are all input channels accepting some subtype of A
  ev1: Tuple.Union[Chans] <:< IChan[A],
  // The match cases are all functions to some process
  ev2: Tuple.Union[Matches] <:< Function1[?, Process],
  // The match cases cover exactly all possible inputs of A
  ev3: Tuple.Union[Tuple.Map[Matches, ArgumentOf]] =:= A,
): ValidBranch[A, Chans, Matches] with {}
\end{lstlisting}

In comparison to match expressions, this implementation provides more control
to the framework authors. We develop a syntax that still feels natural to the
programmer but maintains the necessary type information, especially for
model checking in the future. Whilst every incoming message is guaranteed to
have a corresponding continuation, we unfortunately lose the ability to
detect duplicated labels compared to the formal model.

As well as preserving type information for compile-time checking of implementations,
it is also necessary to preserve the types of the continuation arguments
at \textit{runtime}, so that the evaluator can select the correct continuation
for each incoming message. We achieve this with another implicit
parameter \texttt{WrapMatches}, which converts the tuple of continuations
into a tuple of so-called \texttt{MatchCase} instances. Each instance
gets tagged with a \texttt{TypeTest} for its argument type which we
may use during evaluation.

With the types in place, the DSL can provide a \texttt{branch} function for writing
branching operations, similar to the existing \texttt{send}
and \texttt{receive} functions. This simply instantiates the \texttt{Branch} class.

\begin{example}
    Using the provided \texttt{branch} DSL extension, we may reimplement the
    travel agency from Example \ref{ex:travel-agency-faulty} with
    enforced continuations dependent on the received \texttt{Decision} message.
    
    \begin{lstlisting}[language=Scala]
type TravelAgency[C1 <: IChan[Decision], C2 <: OChan[String]] = 
  Branch1[Decision, C1, (
    (a: Accept) => Out[C2, String], // Provide a ticket
    (r: Reject) => PNil // Do nothing
  )]

def travelAgency(c1: IChan[Decision], c2: OChan[String]):
  TravelAgency[c1.type, c2.type] = {
    branch1(c1, (
      (a: Accept) => send(c2, "Your ticket"),
      (r: Reject) => nil // Sending a ticket here would be a type error
    ))
}
\end{lstlisting}

    \label{ex:branching-dsl}
\end{example}

\paragraph{Timeouts.} To implement the timeout operator, we extend the DSL
in a similar way: by subtyping \texttt{Process} (below). Note the restriction
to \texttt{TimeoutableProcess}, which ensures that the first process provided is
an \texttt{In} or \texttt{Branch} operation. (The \texttt{Out} operation cannot
time out in the current toolkit.)

\begin{lstlisting}[language=Scala]
case class CatchTimeout[P <: TimeoutableProcess, Q <: Process]
  (p: () => P, onTimeout: () => Q) extends Process
\end{lstlisting}

As before, we provide a DSL function to instantiate the class. One notable
property is that the \texttt{CatchTimeout} does not impose a timeout duration
itself. Instead, the duration is specified by the wrapped
\texttt{In}/\texttt{Branch} process.

\begin{example}
    We may now represent the \texttt{AuctionHouse} protocol as originally
    described in Example \ref{ex:auction-house-faulty}, which was not previously
    possible. Using \texttt{CatchTimeout} allows us to respond to a lack of
    incoming bids, and the multi-channel branching allows us to correctly
    handle the auction closing.
    
    \begin{lstlisting}[language=Scala]
type AuctionHouse[C1 <: IChan[Bid], C2 <: OChan[CloseAuction]] =
  CatchTimeout[
    Branch[Bid | CloseAuction, (C1, C2), (
      (b: Bid) => ..., // Handle the bid
      (c: CloseAuction) => PNil // Stop the auction
    )],
    ... // Handle timeout, e.g. reduce starting price
  ]
\end{lstlisting}

    \label{ex:auctionhouse-branching}
\end{example}

\subsection{Evaluating the New Operations}

\effpi\ provides three evaluation strategies: a naive strategy where each
process runs on its own thread; and two more sophisticated strategies where
processes are scheduled cooperatively to run on a fixed number of executor
threads. We briefly describe the modifications to each.

\paragraph{Naive strategy.} This strategy uses a recursive blocking function
\texttt{eval} to interpret the operation tree. We support
branching by polling on each of the channels (in a shuffled order, for fairness) until a
message is available. Supporting timeouts involves setting a
timeout continuation for the next recursive call of \texttt{eval},
which propagates to the evaluation of the wrapped \texttt{In} or \texttt{Branch}
operation. Receiving directly from a channel can specify a timeout duration, and we
catch this exception to trigger the timeout continuation.

\paragraph{Optimised runtime systems.} These systems
share a common executor-based architecture. Readers are referred to the
linked code repository for the full details of the modified implementation.
At a high level, the modifications include generalising channel queues to
support enqueueing both \texttt{In} and \texttt{Branch} operations as waiting
to receive messages. We introduce a timer thread to handle timeouts and
schedule their continuations for execution.

\begin{figure}[h]
    \centering

    \vspace{0.3em}
    \includegraphics[width=0.8\textwidth]{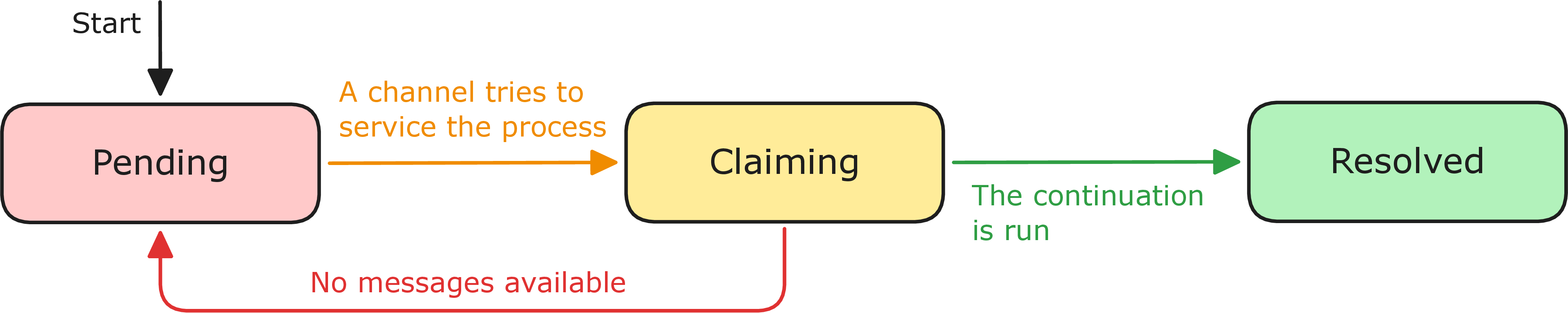}

    \caption{The state of a \texttt{WaitingProcess}, tracked by an atomic variable.}
    \label{fig:waiting-process-state}
\end{figure}

Careful consideration is given to the interaction of
multiple channels attempting to resolve the same \texttt{Branch} operation,
and to races arising between timeouts and incoming messages. It is important
to ensure that only one continuation (whether timeout or message-based) is
triggered for each operation, regardless of the number of channels involved.
We introduce a 3-state atomic variable to track the state of a waiting process:
\texttt{Pending}, \texttt{Claiming}, and \texttt{Resolved}. The intermediate
\texttt{Claiming} state is used to ensure only one channel (or timeout) can
attempt to resolve the operation at a time. When a timeout fires but is
unable to claim the waiting process, it is rescheduled to try again after
a short delay, until it can claim the process or the process is resolved
by an incoming message.
\section{The Raft Election Algorithm}

\subsection{Overview}

Raft \cite{ongaro_raft_2014} is a consensus algorithm for managing a
replicated state machine across a distributed system. It provides guarantees
(such as election safety and leader completeness), even
in the presence of node failures. Raft is widely praised for its
understandability, especially when compared to the Paxos
algorithm \cite{lamport_parliament_1998}, although the two algorithms share
a similar approach to consensus \cite{howard_paxos_2020}.

A key component of the Raft algorithm is the leadership election, which we
explore in detail. Nodes exist in one of three states (\textit{follower},
\textit{candidate}, or \textit{leader}) and transition
between states as internal timeouts expire or as \textit{remote procedure calls}
(RPCs) are exchanged. These have two forms:
\textit{RequestVote} (sent by candidates to gather votes) and \textit{AppendEntries}
(sent by leaders to replicate the log).

Each RPC carries a \textit{term number} which serves as a logical clock and is
incremented with each election. The key property of \textit{election safety}
ensures that there is at most one leader per term.

\begin{figure}[h]
    \centering

    \vspace{0.3em}
    \includegraphics[width=0.8\textwidth]{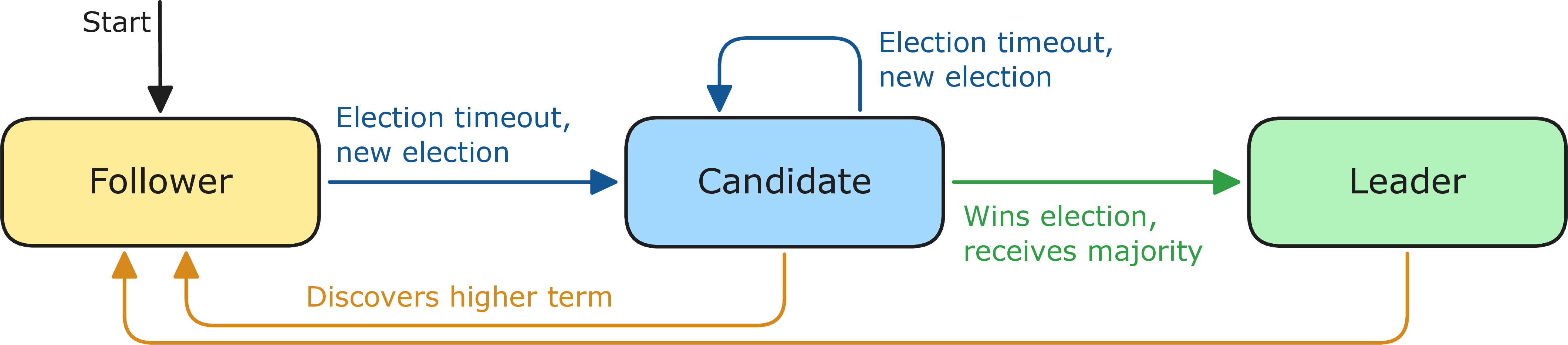}

    \caption{State machine representing Raft node states, adapted from \cite{ongaro_raft_2014}.}
    \label{fig:raft-state-machine}
\end{figure}

We may summarise the election algorithm with respect to the role of each node state.
Additional details, such as the conditions attached to granting votes, have been
omitted for brevity (see \cite{ongaro_raft_2014}).

\begin{itemize}
    \item \textbf{Follower:} Listens for incoming RPCs, replying appropriately.
    If no \textit{AppendEntries} are received for the election timeout period,
    transition to candidate and stand in a new election.

    \item \textbf{Candidate:} Starts a new election, incrementing the term
    number and broadcasting \textit{RequestVote} RPCs to the other nodes.
    If a majority of nodes grant their vote, transition to leader. Otherwise,
    if the election timeout occurs without a majority, start a new election
    and stand for election again.

    \item \textbf{Leader:} Sends periodic \textit{AppendEntries} RPCs to the
    other nodes, acting as heartbeats.
    In the full Raft protocol, the leader
    is additionally responsible for writing new entries to the replicated log.

    \begin{itemize}
        \item In both the \textit{leader} and \textit{candidate} states, if an
        RPC is received with a higher term number, the node transitions back to
        being a \textit{follower}.
    \end{itemize}
\end{itemize}

\subsection{Implementation in Effpi}

With the help of the newly introduced operations, we may
begin to express the Raft election protocol in \effpi\ at the type-level.
For clarity, we approach this task in several stages.

\paragraph{Timer process.}
The protocol uses timeouts to trigger
state transitions: for example, followers become candidates if they do not
receive heartbeats within a randomly chosen election timeout period. To model
this behaviour, we define a \texttt{timer} process which is responsible for
producing timeout events and handling reset instructions from the main node
process. We model the timer as a separate process to allow for more
fine-grained control over when to reset the timer and listen for timeout events.
Many of Raft's receiving/branching operations should only reset the timer on
some of their continuations, for example. 

\begin{lstlisting}[language=Scala]
type Timer = Rec[RecX,
  In[timerReset.type, TimerReset, TimerReset => Rec[RecY,
    CatchTimeout[
      // If we receive a timer reset before timeout, we loop to RecY.
      In[timerReset.type, TimerReset, TimerReset => Loop[RecY]],
      // Once we time out, we loop to RecX: no pending timeout!
      Out[timeoutChan.type, TimerExpired] >>: Loop[RecX]
    ]]]]
\end{lstlisting}

By using two recursive variables, \texttt{RecX} and \texttt{RecY}, we model
the different behaviours of the timer process depending on whether it has
received a reset instruction since the last timeout event. After receiving
the first \texttt{TimerReset}, we use the new \texttt{CatchTimeout} operator:
we accept reset messages, but produce a \texttt{TimerExpired} message if no
reset message is received in time.

\paragraph{RPC reply behaviours.}
Before defining the processes for each node state, we define common
behaviours shared between them. Replying to a \textit{RequestVote} RPC involves
deciding to grant or deny the vote and performing the corresponding
actions. We encapsulate this in \texttt{GrantVoteBehaviour} and
its \texttt{Deny} counterpart, and call their union
\texttt{VoteReplyBehaviour}. The type parameter \texttt{C} represents the
channel to reply to, ensuring the correct channel is used.
After granting a vote, the node is always expected to become a follower,
whereas after denying a vote, it should return to its current state.
We capture this via another type parameter \texttt{V}, representing the recursion
to loop back to.

\begin{lstlisting}[language=Scala]
type GrantVoteBehaviour[C <: OChan[VoteResponse]] =
  Out[C, GrantVote] >>: Out[timerResetChan.type, TimerReset]
    >>: Loop[RecFollower]

type RefuseVoteBehaviour[C <: OChan[VoteResponse], V[A] <: RecVar[A]] =
  Out[C, RefuseVote] >>: Loop[V]
\end{lstlisting}

\paragraph{Follower state.}
We now define the full Raft election protocol state by state, beginning
with followers. At its core, we define the process as
repeatedly receiving messages over the inbox and timeout channels,
replying to the RPCs appropriately. Dependent function types \cite{scala_dependent_function_types}
allow us to extract the type of the reply channel from the incoming message. Passing
the reply channel type as a parameter to the reply behaviours ensures the
correct channel is used for replies. Upon receiving a timeout, we transition
to become a candidate. The \texttt{branch} operation is instrumental in
enforcing the correct continuations are run and enables us to listen on both an
external inbox channel and the internal timeout channel.

\begin{lstlisting}[language=Scala]
type Follower = Rec[RecFollower,
  Branch[RpcMessage | TimerExpired, (inbox.type, timeoutChan.type), (
    (rv: RequestVote[_]) => VoteReplyBehaviour[rv.reply.type, RecFollower],
    (ae: AppendEntries[_]) => AEReplyBehaviour[ae.reply.type, RecFollower],
    TimerExpired => Candidate
  )]]
\end{lstlisting}

From this type-level description, we may implement the process straightforwardly,
simply using the DSL and filling in details such as
the logic for deciding between granting and denying votes. The definition of
the protocol as a type constrains the process definition, reducing possible
implementation errors. Readers are
referred to the open source code repository for the full program.

\paragraph{Candidate state.}
In addition to replying to incoming RPCs like followers do, candidates must also
initiate an election, broadcasting \textit{RequestVote} messages to
the other nodes. We define the candidate in two parts: first, a
\texttt{CandidateElection} type which handles the lifetime of a single
election; and second, the full \texttt{Candidate} type which starts a new
election. Splitting the definition in this way allows us to capture the fresh
reply channel as type parameter \texttt{C}. This forces the broadcasted
\textit{RequestVote} messages to specify the same channel as the one
we listen on for incoming vote responses.

\begin{lstlisting}[language=Scala]
type CandidateElection[C <: Chan[RequestVote]] = Par[
  Broadcast[RequestVote[C]], // Broadcast vote requests, with replyChan = C
  Rec[RecCandidate,
    Branch[RPC | TimerExpired | VoteResponse, (C, /* channels */), (
      ... // Handle the incoming RPCs and vote responses
    )]
  ]]

type Candidate = Rec[RecElection, Out[timerReset.type, TimerReset]
  >>: CandidateElection[Chan[VoteResponse]]]
\end{lstlisting}

\paragraph{Leader state.}
Finally, a leader process may be defined, following the same
design principles. A key difference is that the leader must broadcast
\textit{AppendEntries} RPCs periodically. This is achieved by making use
of the timer process and using two recursive variables, much like the candidate
state. The full implementation is again deferred to the code repository.

\begin{lstlisting}[language=Scala]
type Leader = Rec[RecLeaderHeartbeat, // First recursive variable
  Out[timerReset.type, TimerReset]
    >>: Par[
      Broadcast[AppendEntries[Chan[AckAppendEntries]]],
      Rec[RecLeader, // Second recursive variable
        Branch[RPC | TimerExpired, (inbox.type, timeoutChan.type), (
          (ae: AppendEntries[_]) => ..., // As before
          (rv: RequestVote[_]) => ..., // As before
          TimerExpired => Loop[RecLeaderHeartbeat]
        )]
      ]
    ]]
\end{lstlisting}

\paragraph{Discussion.}
By extending the \effpi\ toolkit, we have been able to better express the
Raft election algorithm at the type-level, capturing the required behaviours
more faithfully. The implementation of each process is type-checked against
the defined protocol type, making it easier to identify errors now that
branching and timeout operations are enforced by the type system.
However, there are still a number of limitations, such as the lack of
model checking support for programs using the new operations. Creating a
fresh channel (as in \texttt{CandidateElection}) could also be better
supported. Note we restrict Raft to a fixed number of nodes: the toolkit could
natively support any number of nodes by introducing a true broadcasting
operation.

\section{Conclusion \& Related Work}

In this work, we presented an extension to the \effpi\ toolkit, by means
of the \extendedLambdaPi\ calculus and concrete Scala
implementation. The new branching and timeout operations improve the
expressiveness of \effpi, which we demonstrated through two smaller examples
and an implementation of the Raft election algorithm. The enhanced expressiveness
enables more faithful representations of protocols as types, allowing us to
encode more precise protocol specifications and thereby
better ensure process implementations adhere to their protocols.

\paragraph{Related work.}
Effpi was introduced by Scalas \etal\ \cite{scalas_effpi_2019,scalas_verifying_2019}
and has since been used for code generation from global Scribble types \cite{barwell_crash_2023}.
Few other extensions have been proposed to the toolkit to date. Frameworks such as Akka Typed
\cite{lightbend_akka_2019} and \texttt{lchannels} \cite{scalas_lightweight_2016}
are alternative toolkits for writing typed message-passing programs in Scala,
the latter using session types as the underlying type system.

External choice is a common operation in concurrency theories.
Multiparty session types \cite{yoshida_less_2024} include branching and
selection as core constructs; unlike our approach, 
there are no separate send/receive operations, only branch/select.
This shows how the branching operation could be used as a replacement
for the existing receive operation, rather than as an extension.
The $\pi$-calculus implements choice through
a ``$+$'' operator, which offers input on possibly multiple channels
\cite{milner_pi_1999}. Communicating Sequential Processes (CSP) also
includes external choice \cite{hoare_csp_1978}.

Timeouts have been studied in a number of works. Notably, CSP includes a
sliding choice operator \cite{roscoe_tpc_2000}, which inspired our own
operation. Timed extensions to the $\pi$-calculus have been proposed
\cite{timed_pi_2013} as well as extensions to multiparty session types
\cite{ping_fearless_2024}, both of which handle time constraints
more generally.

\paragraph{Future work.}
Future work includes proving the formal properties of the extended type system
and implementing model checking support for the new constructs
in the compiler plugin. This will enable the verification of deadlock-freedom,
among other properties, of programs that use branching or timeouts. We aim to verify
the Raft implementation presented here and explore the use of the model checking
to prove properties such as election safety. Raft has previously
been modelled in mCRL2 \cite{bora_raft_2024}, which may provide a useful
comparison for this verification.

\paragraph{Acknowledgements.}
The authors would like to thank Dylan McDermott and the PLACES reviewers
for their helpful comments.
The first author is partially supported by a travel grant from St Hugh's
College, University of Oxford.
The second author is partially supported by EPSRC grants
EP/T006544/2, EP/T014709/2, EP/Y005244/1, EP/V000462/1, EP/X015955/1, EP/Z0005801/1;
Horizon EU TaRDIS 101093006 (UKRI No. 10066667); and ARIA. 

\bibliographystyle{eptcs}
\bibliography{references}

% FOR LONG VERSION ONLY:
% \clearpage
% \input{appendix}

\end{document}